\documentclass[useAMS,usenatbib]{mn2e}

\usepackage{graphicx}

\def\lta{\mathrel{\spose{\lower 3pt\hbox{$\mathchar"218$}}
     \raise 2.0pt\hbox{$\mathchar"13C$}}}
\def\gta{\mathrel{\spose{\lower 3pt\hbox{$\mathchar"218$}}
     \raise 2.0pt\hbox{$\mathchar"13E$}}}
\newcommand{\etal}{{\rm et al. }}

\def\mathnew{\mathsurround=0pt}

\def\simov#1#2{\lower .5pt\vbox{\baselineskip0pt \lineskip-.5pt
\ialign{$\mathnew#1\hfil##\hfil$\crcr#2\crcr\sim\crcr}}}

\def\simless{\mathrel{\mathpalette\simov <}}

\title[Incidence of KHz QPO in the neutron star system 4U 1636-53]{On the incidence of KHz quasi-periodic oscillations in the neutron star system 4U 1636-53}
\author[R. Misra and  K. Shanthi]{R. Misra$^{1}$\thanks{E-mail:
rmisra@iucaa.ernet.in} and K.
Shanthi$^{2}$\thanks{E-mail:
kshanti@mu.ac.in}\\
$^{1}$Inter-University Centre for Astronomy and Astrophysics, Post Bag 4, Ganeshkhind, Pune-411007, India\\
$^{2}$UGC Academic Staff College, University of Mumbai, Mumbai-400098, India}
\begin{document}

\date{Accepted - . Received -; in original form }

\pagerange{\pageref{firstpage}--\pageref{lastpage}} \pubyear{2004}

\maketitle

\label{firstpage}

\begin{abstract}
Nearly 
$1.32 \times 10^6$ sec of non-zero count rate data were obtained from
 five years of {\rm RXTE} observations of the 
atoll source 4U 1636-53.
The data was
divided into 10309 segments of 128 sec each. The histogram of the
number of segments as a function of count rate shows that the
system can naturally be classified into four flux states. For each segment 
an automated search for kHz QPO was undertaken and
the histograms of the maximum Leahy normalised power $P_{max}$ 
for the four flux states were created. These were fitted by 
probability distribution functions under the assumption that the 
intrinsic amplitude of the QPO
has a normal distribution. The best fit distribution functions,
showed that the fraction $f$ of the 
time the system exhibits a kHz QPO, 
decreases from near unity for the lowest
flux state ($\approx 1400$ c/s) to zero for the highest one
($> 2550$ c/s) while the average amplitude remains nearly constant.
Based on the best fit probability distribution functions, 
a threshold $P_{thre}$ was defined, such that 95\% of the segments
with $P_{max} > P_{thre}$ correspond to a real QPO signal. For
segments selected by this criterion, the
frequency versus count rate plot reproduced the earlier known ``parallel
track'' variation with the tracks coinciding roughly with the flux states.
It is shown that the gaps between the tracks are not caused by uneven sampling,
but rather the QPO phenomenon is absent or weak when the system's flux level
is  intermediate between two flux states. 
\end{abstract}


\section{Introduction}

Rapid (300--1300~Hz), nearly periodic variability in the X-ray light curves 
of low mass X-ray binary systems (LMXB's) have been observed by
the Rossi X-ray Timing Explorer (RXTE) in over two dozen neutron 
star bearing LMXB's (see van~der~Klis~2000 for a review).
These oscillations, referred to as kilohertz QPO 
(quasi-periodic oscillations), often have high quality factors 
(Q=FWHM/frequency), and tend to 
be seen in pairs, with nearly constant frequency separation between 
the two peaks.
The high frequencies of kilohertz QPO strongly suggests that
they are related to phenomena 
taking place in the inner regions of accretion disks surrounding neutron stars,
making them potentially valuable probes of strong gravity and the behaviour
of matter in such environments.

Several theoretical ideas have been proposed to explain the phenomenology 
of kilohertz QPO.  In these models, one of the frequencies is generally 
identified 
as the Keplerian frequency of a characteristic radius of an accretion disk
(for e.g. the innermost orbit). In the sonic point
model (Miller, Lamb \& Psaltis 1998) the second frequency is identified as 
the beat of the primary QPO with the spin of the neutron star. 
\cite{ST99} have proposed a general relativistic precession/apsidal motion model 
wherein the secondary QPO is due to the relativistic apsidal motion of this
characteristic orbit.  On the other 
hand, in the two oscillator model \cite{OT99}, the secondary 
frequency is due to the transformation of the primary (Keplerian) frequency 
in the 
rotating frame of the neutron star magnetosphere. Recently \cite{L04}
have proposed that the QPO occurs at that radius of the disk where  local
resonance
occurs  between the Keplerian and epicyclic frequencies.
These models are based on their predictions for the  
variation of the frequency 
separation with the kilohertz QPO frequency and the variation of 
the kHz QPO frequency with that of  
other QPO observed in a source.  
  However a clear consensus regarding which is the viable model has
not arisen perhaps because  most of these 
dynamic models generally do not address the direct origin
of these oscillations. 
Using the observed lags in different
energy bands, Lee, Misra \& Taam (2001), showed that the QPO is driven by a temperature
variation in a Comptonizing cloud of size $\approx 10$ km.  
A unified model for kilohertz QPO, that takes into account both the
dynamic and radiative aspects of the phenomenon is still illusive.
Such a model should 
also address how often and under what circumstances does  a QPO
occur and hence the primary motivation of this work is to quantify the  
incidence of the kHz QPO phenomenon in the atoll source
4U 1636-53. It is important to know whether the phenomena
is a ubiquitous or a transient one, which in turn will provide insight into
its cause.

In the next section, we argue that the standard data analysis technique
is not adequate to address this question and motivate the
need for a modified technique. In that section we highlight the reasons why 
the atoll source  4U 1636-53 was chosen for the analysis and the
need to study the incidence in terms of the long term flux states.
In \S 3 the
data analysis  technique and the theoretical probability distribution
functions which are fitted  to the results of the analysis are
described. In \S 4 the results
are presented and fitted while in \S 5 the implications of the results are  
discussed.

\section{Data analysis Technique}

The standard technique to analyse the kHz QPO phenomenon is
to  compute  power spectra of segments of the light curve
and fit Lorentzian profiles to them. This provides 
unambiguous measure of the QPO frequency, width and amplitude.
Correlation between these quantities and the spectral properties
like intensity, hardness ratio can be tested and quantified.
However, it is difficult to use this technique to obtain
an estimate of the incidence of the QPO phenomenon. To
apply this technique, one has to impose a threshold significance
level, say $95$\% confidence level, that a QPO has been detected
for a given segment of data. This means,  by definition, that
a certain fraction ($5$\% in this example) of the segments
analysed will have positive but false detections. These detections
may be a significant fraction of the real QPO that can be
detected in this scheme ( say $10$\%). Increasing the
threshold level, will decrease the number of QPO detected and
biases the result toward high amplitude detections. 
Another related problem with this analysis technique is the
bias in the measure of the amplitude. To illustrate this, consider
a long light curve which has a QPO with a constant amplitude $A$,
which is less than the threshold significance level of the analysis.
The resultant power in each segment due to this QPO will not be
a constant but will have statistical fluctuations. Hence, occasionally,
the power will be greater than the threshold and this result can be
interpreted as being due to a rarer but stronger QPO in the light curve.
As noted by \cite{leahy}, significance of the presence and amplitude 
of a detected QPO, should take into account the number of segments
and frequency range searched. The results obtained without taking
these effects into account maybe biased and should be interpreted
with caution.

These deficiencies of the standard data analysis technique
are difficult to overcome and hence to make progress, significant
trade off in the accuracy and interpretation of the data have
to be made. The technique applied in this work, which is
described in more detail below, involves finding the Leahy
normalised power spectrum for each segment and identifying
the maximum power and the frequency at which that occurs. The histogram
of the maximum power can then be compared with theoretical probability
distributions.
The  advantage of this method is that there are no thresholds for
QPO detections imposed and hence the data is used maximally. However, there
are two related drawbacks. First, the peak of the power spectrum is
representative of the peak of the Lorentzian fit to the QPO
and not the amplitude. To
obtain the amplitude one needs to know also the width 
or Quality factor of the QPO, which is not extracted in this
analysis. Second, the theoretical probability distribution,
in principle, should also depend on this width or quality factor.
Moreover, these probability distributions
for Leahy normalised maximum power are computed on the assumption that the
the count rate is a constant for different segments, which is not
true even when the data is selected for a given intensity state.
Taking these effects correctly into account is 
complicated and hence in this
work we make simplistic assumptions when computing the
theoretical probability distributions and partially justify
our choice by showing that they fit the observed distributions
with a reduced $\chi^2$ or order unity. This means that more
correct but complicated probability distribution functions, perhaps cannot
be differentiated by the present data and hence may not be warranted.

The presence and properties of the kHz QPO are expected to be
correlated to the source's spectral parameters (see 
van~der~Klis~2000 for a review and references therein). Low mass X-ray
binaries containing weakly magnetised neutron stars are
divided into two classes based on the Z or atoll like shaped
tracks they trace out in colour-colour diagrams. The time-scale
to trace out the tracks is typically short ($\simless$ day),
although atoll sources can take longer ($\simless$ week) to
cover the entire curve.
The variation in the intensity
during such a trace out is typically a factor  $\simless 2$, while
the changes in the colours (depending on the definition) is
typically $\simless 20$\%.  The kHz QPO properties are
known to be correlated to the position of the source in
the colour-colour diagram in these timescales. Over longer time-scales
the behaviour of the system is more complex. For some of the atoll sources,
there are secular variation in the intensity by $\approx 20$\% while
the response corrected colour-colour track remains nearly invariant 
(Di Salvo, Mendez \&  van~der~Klis 2003).
The QPO frequency versus intensity plots show parallel tracks
where each track marks the short term correlation between the two
(e.g. Mendez \etal 1999 ; Mendez, van~der~Klis  \& Ford 2001).
This perhaps implies that the properties of the QPO are determined
by the spectral shape (i.e. the position in the colour-colour diagram)
and not on the long term intensity variations. This is partially
supported by the plot of QPO frequency versus position in the
colour-colour plot, where the parallel tracks in the intensity
plot almost overlap each other \citep{smv00}. 
However, the scatter of the frequencies
with respect to colour-colour position is larger than the scatter
along each track in the intensity plot. Moreover, for the long term
analysis there is no significant correlation between the frequency
and colour-colour position \citep{smv00}. Evidence that the QPO properties
depends on the long term intensity, comes from the frequency-intensity
parallel tracks for different sources, where for all of them the
low intensity tracks have a wider range of observed frequencies
compared to the high intensity tracks. As described in this
work, the histogram of the
number of detections as a function of intensity shows peaks which
nearly coincide with the parallel tracks, which implies that
the source has distinguishable intensity (or flux) states.
In summary, it is not clear if all of the QPO properties depend
only on the spectral shape. The long term intensity may also
be an important factor determining some properties 
especially the incidence.

In this work we quantify the
incidence of the QPO phenomenon as a function of different
long term intensity states. This source was chosen since
a highly coherent kHz ($\approx 800$ Hz)  QPO has been observed in its power
spectrum \cite{Z96,W97}. The second QPO at $\approx 1200$ Hz is weak
and in general cannot be detected by the techniques used in this
work.

\section{Data analysis and Probability distribution functions}

Nearly all of the archival {\it RXTE} data for the atoll source
4U 1636-52, that had the Event mode data (E\_125$\mu$s\_64M\_0\_1s) 
with $128 \mu$sec
time resolution, were analysed.   The resulting $1.32 \times 10^6$ sec
of non zero count data were divided into $10309$ segments of $128$ sec each.
For each segment a power spectrum was obtained at a Nyquist frequency
of $4096$ Hz, by averaging $256$ sub-segments of $0.5$ sec each. Thus
the power spectra had 2048 frequency bins of width 2 Hz. The
power spectrum were Leahy normalised \citep{leahy} i.e.
\begin{equation}
P(f)  = 2 | a(f) |^2 / N
\end{equation}
where $a(f)$ is the Fast Fourier Transform and $N$ is the
total number of counts.   
For each power spectrum the maximum power $P_{max}$ within the
frequency range $200$ to $2000$ Hz and the corresponding frequency $f_Q$
were obtained.
For each segment we define a normalised count rate
$R_N \equiv R (5/n_{pon})$ where $R$ is the detected count rate and
$n_{pon}$ is the number of proportional counter units (PCU) on during the
observation. It should be noted that since in the computation of
$R_N$, the time dependent detector response is not taken into account,
it is at best an approximate representation of the real flux from
the source.

In the absence of a signal, the average power at any frequency 
is $<P> = 2$ and the distribution of $P$, $p_{ns} (P)$ is the $\chi^2$ 
distribution with $2M$ degrees
of freedom, where $M = 256$, is the number of sub-segments \cite{leahy}.
The probability density of obtaining a  
$P_{max}$ is 
\begin{equation}
{\rm P}_{ns} (P_{max},N_f) = N_f\; p_{ns}(P_{max})Q_{ns}(<P_{max})^{N_f-1}
\end{equation}
where $N_f = 900$ is the number of frequency bins sampled and
\begin{equation}
Q_{ns} (< P_{max}) = \int_0^{P_{max}} p_{ns} (P) dP
\end{equation}
is the probability that the power in a frequency bin is less than
$P_{max}$. In the presence of a signal the computation of
the probability density of $P_{max}$, ${\rm P}_s (P_{max})$ is more
complicated and model dependent. For a pure sinusoidal
signal, the average power at the oscillation frequency is
\cite{leahy}
\begin{equation}
<P >  = 2[1+0.19 N A^2]
\end{equation}
where $A$ is the amplitude of the oscillation. It has been
assumed here that the oscillation frequency is much smaller 
than the Nyquist one.
The probability distribution  $p_s (P,N,A)$ can be approximated to be a 
$\chi^2$ distribution of 2M degrees of freedom but rescaled such that
the average power is given by eqn (4). 
The total number of
counts $N$ is not a constant even for all segments considered in the same
flux state, since the number of PCU that
are on during a observation varies. This variation can be taken into
account by representing the probability distribution as
\begin{equation}
p_s (P,A)  = \sum_{i = 1}^5 g_i p_i (P,N_i, A)
\end{equation}
where $i$ is the number of PCU on, $g(i)$ is the fraction
of the time that $i$ PCU were on and $N_i$ is the
average number of counts when $i$ PCU were on i.e
$N_i = R_N (n_{pon}/5) \times 128$ sec.
The above expression 
assumes that the amplitude of the oscillation is a constant
which naturally may not be the case. Moreover, in a defined flux
state, the value of $R_N$ varies over a range which will affect
the predicted average power (eqn 4). Further, the observed
QPO is not a pure sinusoidal signal and has a finite width. 
Taking these effects into account is complicated and hence for
simplicity 
we convolute the probability distribution $p_s (P, A)$ with
a Gaussian of width $\sigma$ to give a averaged probability distribution
$<p_s(P,A)>$. 
The probability density for obtaining
$P_{max}$ under these assumption becomes
\begin{eqnarray}
{\rm P}_s (P_{max},\sigma, A)& = & {\rm P}_{ns} (P_{max},N_f-1) Q_s (< P_{max}) + \nonumber \\ 
& & <p_s (P_{max},A)> Q_{ns} (< P_{max})^{N_f-1} 
\end{eqnarray}
where 
\begin{equation}
Q_{s} (< P_{max}) = \int_0^{P_{max}} <p_s (P,A)>  dP
\end{equation}
is the probability that the power in the oscillation frequency bin is less than
$P_{max}$. The first term in eqn (6) is the probability density that
$P_{max}$ is obtained by chance at any other frequency, while the
second term is for the probability of $P_{max}$ to occur at the
oscillation frequency.
At a given flux state,
in general, a signal may exist only for a fraction $f$ of the time.
Hence the total probability distribution is
\begin{equation}
{\rm P} = f {\rm P}_s (P_{max},\sigma, A)  + (1-f) {\rm P}_{ns} (P_{max},N_f)
\end{equation}
The probability distribution ${\rm P}$ has three unknown
parameters $f$, $A$ and $\sigma$ and depends on the known
normalised flux $R_N$ and PCU on fractions $g_i$. 

The technique used in this analysis is sensitive only to narrow band,
peaked signals. Any broad band component will be spread over all
(or a significant fraction of) the 900 frequency bins used. Thus the
contribution of a equally strong broad band signal at a given frequency bin 
would be a large factor ($\approx 100$) less than that of a narrow band signal 
which is spread over fewer ($\approx 10$) frequency bins. 
Thus, to effect the analysis a broad band signal would need to 
have at least $10-100$ time more power than a typical QPO detected 
($\approx 5$\%). For an average rate of 2000 counts/sec,
the Poissonian noise at 500 Hz ( 300 Hz)  is 100\% (30\%) and 
hence the Poissonian noise is expected to dominate over any underlying
broad component which justifies the use of $\chi^2$ distribution for
$p_{ns} (P)$. However, it should be noted that the effect of any such
high frequency broad band noise \citep[e.g.][]{Str00} has not been
taken into account in this analysis.

\section{Results}

The histogram of the  normalised count rate $R_N$ for 
all the data is shown in Figure 1. $R_N$ ranged from
$\approx 500$ counts/sec to $\approx 4000$ counts/sec, not
including the burst periods. We have classified the data into
four flux states which are marked in Figure 1. Data for $R_N < 1100$
counts/sec have not been shown in the figure, since
the low count rate and the relatively small number of detections
does not allow for any significant results to be obtained as discussed
below.
Figure 2 shows the histograms of $P_{max}$ for the four states.
The error for each bin was taken to be the square root of the 
number of detections at that bin. Fitting was done
over a continuous range of $P_{max}$ where there were detections.
The dotted lines represent the corresponding curves if there
were no signals in the data ${\rm P}_{ns} (P_{max})$.
Note that there are no free parameters in ${\rm P}_{ns} (P_{max})$ 
and the good fit to the high state (IV) histogram (reduced $\chi^2 = 0.6$)
 implies that 
there are no strong oscillatory signals in this state and
the fraction of time a QPO is present during this state
is nearly zero. The upper limit on this fraction $f < 0.06$ for $A > 0.025 $
can be inferred.
The shape of the
histograms for the other three states are clearly different,
indicating that there are
oscillatory signals at least for some fraction of the time during
these states. These histograms were fitted by the 
probability distribution ${\rm P}$ for different values of  fraction $f$, 
and amplitude $A$ while $\sigma = 0.15$ was held constant. 
The best fit curves (solid lines), the best fit parameter values and the number of
degrees of freedom are shown in in Figure 2. 
Varying $\sigma$ does not
change the best fit parameters significantly. For example,
for $\sigma = 0.25$ the best fit values of $f$ are
$0.97$, $0.36$ and $0.15$ for states I,II and III respectively
while the corresponding values for $\sigma = 0.05$ are
$0.8$, $0.30$ and $0.11$.
For $R_N < 1100$ counts/sec, the best fit parameters obtained
were $f = 1.0^{+0}_{-0.4}$ and $A =  .0475\pm 0.005$ which are
consistent with the interpretation that $f \approx 1$ for low count rates.
The theoretical probability distributions adequately represent
the data with reduced $\chi^2 \approx 1$ for all the fits. Hence
more sophisticated probability functions are perhaps not warranted
and may lead to over modelling.

\begin{figure}
\begin{center}
{\includegraphics[width=0.8\linewidth,angle=0]{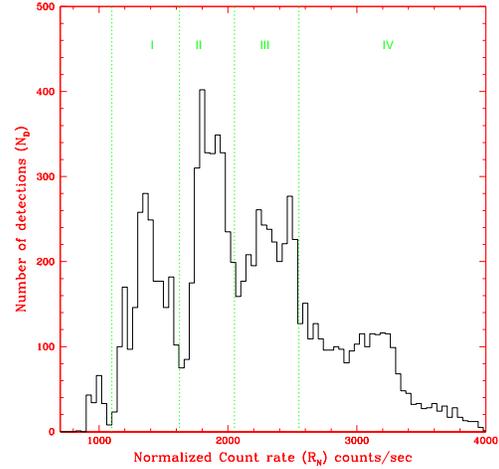}}
\end{center}  
\caption{ The histogram of normalised count rate $R_N$ for all the
data sets considered. Four flux states have been identified and marked
by roman numerals. I: $ 1225 < R_N < 1625 $
counts/sec, II: $ 1625 < R_N < 2050 $ counts/sec, III: $ 2050 < R_N < 2550 $
counts/sec, IV: $R_N > 2550$ counts/sec.\label{Fig. 1}
}
\end{figure}

\begin{figure}
\begin{center}
{\includegraphics[width=0.8\linewidth,angle=0]{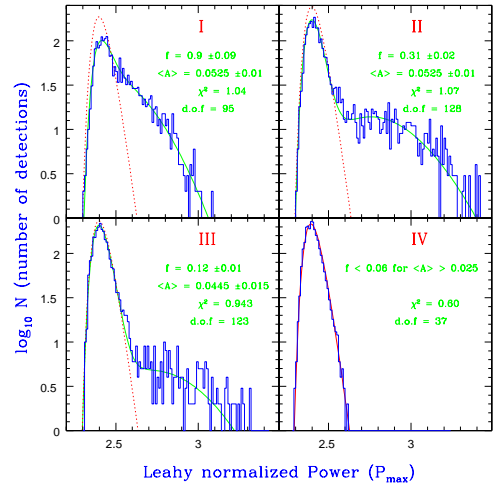}}
\end{center}  
\caption{ The histograms of the maximum Leahy normalised power
$P_{max}$ for four flux states. The dotted line is the 
expected probability distribution due to statistical fluctuations 
alone ${\rm P}_{ns} (P_{max})$. The best fit parameters, the
reduced $\chi^2$ and the number of degrees of freedom, d.o.f, are given. 
For all probability distributions $\sigma = 0.15$.
\label{Fig. 2} 
}
\end{figure}

Based on the best fit probability distribution,
We can now define a threshold power $P_{thre}$ such that for 95\% of
the segments with $P > P_{thre}$, the peak power corresponds to a real
signal. The corresponding thresholds for the states, I, II and III turn out
to be $2.55$, $2.56$ and $2.60$ respectively.  
The completeness of this selection criterion i.e.
the fraction of the total number of segments with QPO that would
be selected turn out to be $0.36$, $0.82$ and $0.69$ 
for the three states respectively.
The corresponding frequencies of the segments with $P > P_{max}$ are
shown as a function of the normalised count rate in Figure 3. 
As expected, the  parallel tracks are reproduced. The histogram
of the selected segments for different count rates (Figure 4 top panel) shows
peaks at the same fluxes as in  Figure 1. This may indicate that
the gaps in the frequency-flux plot (Figure 3) maybe due to
the uneven distribution of fluxes exhibited by this source. However,
the ratio of the segments with $P > P_{thre}$ to the total number
of segments at a flux level (Figure 4 b) also shows peaks at the
same corresponding fluxes. Thus it appears that there are at least
three distinct flux levels for this source. The QPO phenomenon appears
when the system is in one of these flux level , while in the
intermediate flux levels the phenomenon is either weak or absent.

\begin{figure}
\begin{center}
{\includegraphics[width=0.8\linewidth,angle=0]{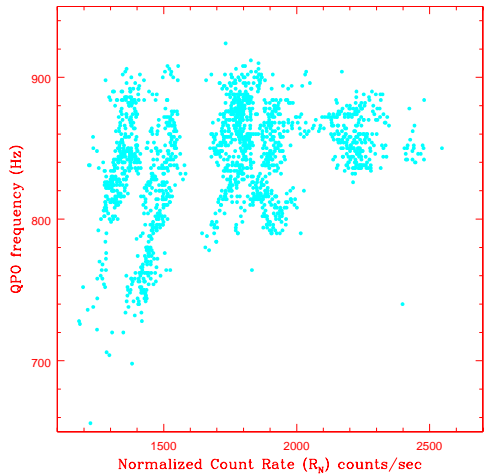}}
\end{center}  
\caption{The variation of the QPO frequency for all segments
with $P > P_{thre}$ versus the normalised count rate.\label{Fig. 3} 
}
\end{figure}

\begin{figure}
\begin{center}
{\includegraphics[width=0.8\linewidth,angle=0]{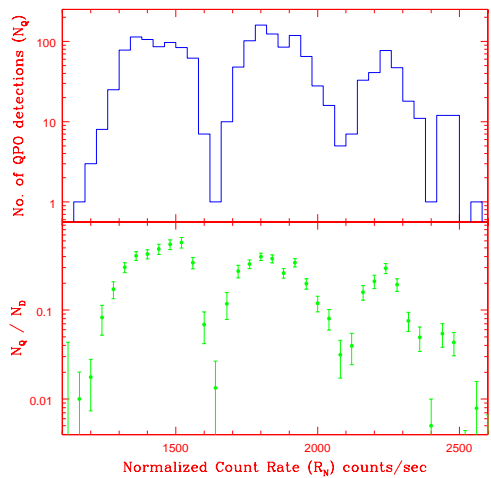}}
\end{center}  
\caption{ The number of segments  with $P > P_{thre}$, $N_Q$ (Top
panel) and the ratio of $N_Q$ to the total number of segments, $N_D$ (Bottom
panel) as a function of the normalised count rate ($R_N$).
\label{Fig. 4}
}
\end{figure}

The histogram of the number of segments with $P > P_{thre}$ as a
function of frequency is shown in figure 5, while figure 6  shows
the average 
peak amplitude $<A_Q>$, defined as $A_Q \equiv (P - 2)/(0.38 N)$. 
Although, $A_Q$ is a measure of the peak of the Lorentzian function,
rather than the actual amplitude of the QPO, the results obtained
here are similar to the result obtained by \cite{M01,smv00}, who found that
the amplitude is nearly constant till $f \approx 850$ Hz and then
falls off. However, as mentioned earlier
these results should be treated with caution since the selection
criterion makes the QPO sample incomplete.

For a selection criterion based on flux levels, as has been done in
this work, the QPO detected may arise from different parallel tracks.
In principle, each track may have a different average amplitude and/or
incidence. Such a dichotomy may give rise to a bimodal
distribution of QPO powers and should be fitted by two probability
distribution functions. This may be particularly true, when the frequency 
shift between the two tracks is large and the flux range chosen to define 
a flux level is small. However, the flux range chosen in this work, is
sufficiently large such that the corresponding range of frequencies for
each track is large with significant overlap. Thus, the distribution
of QPO versus frequency (Figure 5) does not show any bimodal structure
for states (II) and (III). For state (I) there may be a double peak structure
in the distribution. However, for this state the  average amplitude
does not seem to vary much with frequency (Figure 6) implying perhaps
that the average amplitude is nearly same for the two tracks. Fitting
the detection distribution versus maximum power (Figure 2; top left panel) 
with two probability functions, instead of one, does not lead to a
reduction in $\chi^2$. Instead a degenerate fit where the amplitudes
of the two probability distributions are nearly equal and the 
incidence fractions add up to $\approx 0.9$, is obtained.  
Any difference between the incidence and average amplitude of the
two parallel tracks may be detectable with the use of smaller flux ranges, 
but this leads poor statistics due  to the lower number of detections 
for each flux level. Alternatively, this may be achieved, by a different
selection criteria which uses some other source property  that
can clearly distinguish and separate out the different parallel tracks. 
However, it is not clear which source property can be used here, since the QPO 
frequency has a weak dependence on hardness ratios \citep{smv00}.

\begin{figure}
\begin{center}
{\includegraphics[width=0.8\linewidth,angle=0]{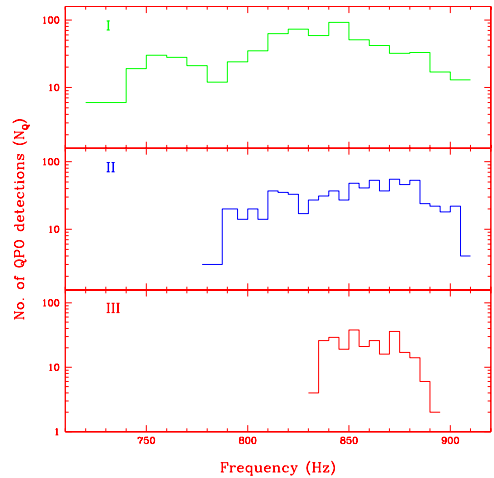}}
\end{center}  
\caption{  The number of segments  with $P > P_{thre}$, $N_Q$  
versus the QPO frequency for the three lower flux levels (I,II and III). 
\label{Fig. 5}
}
\end{figure}

\begin{figure}
\begin{center}
{\includegraphics[width=0.8\linewidth,angle=0]{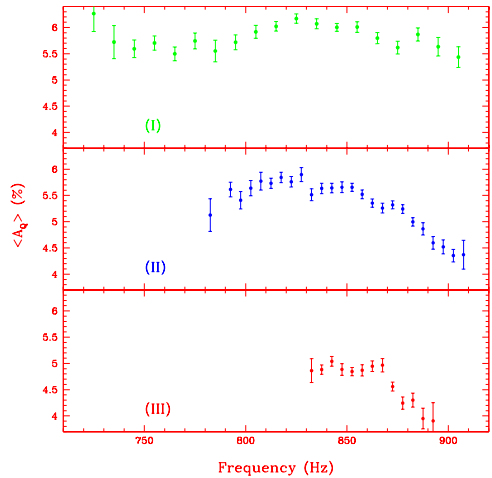}}
\end{center}  
\caption{ The averaged amplitude $A_Q$ for segments with  with $P > P_{thre}$,
versus the QPO frequency for the three lower flux levels (I,II and III).
  \label{Fig. 6} 
}
\end{figure}

\section{Summary and Discussion}

Based on the histogram of the number of detections as a function of
normalised count rate, four distinct flux levels have been
defined. It is found that the incidence of kHz QPO, decreases
from near unity at low flux levels to nearly zero at high ones.
The variation of frequency with count rates reveal roughly three distinct
parallel tracks which are coincident with the lower three flux levels.
It is however, argued that in the intermediate flux levels, the
QPO phenomenon actually disappears or weakens and the parallel
tracks are not due to uneven sampling of the source at different
intensities. 

This analysis suggests that the QPO phenomenon is also related
to the long term flux level of the source and in particular
occurs more frequently as the flux level decreases. The existence
of an inferred threshold accretion rate, beyond which the QPO
phenomenon disappears has been suggested earlier \citep{C00} based
on short term evolution of the source. Here
we show that for the long term evolution there exists such a
threshold intensity level. Within each long term flux level, it
is shown that the QPO weakens and/or disappears as the count rate
increases. Thus the phenomenology seems to be more complex than
envisioned before. \cite{C00} have proposed that the QPO is produced
as a result of the interaction of the magnetosphere with the Keplerian
disk, and at high accretion rates, when the magnetosphere
radius is smaller than the last stable orbit, the phenomenon disappears.
If the magnetosphere radius is the radius at which magnetic pressure
equals the ram pressure of freely accreting material, then this
model predicts a single accretion rate threshold above which
the phenomenon should disappear. Thus this simple model needs
to be modified to explain the disappearance or weakening of the QPO
phenomenon for each long term flux level. This may not be surprising,
because the magnetosphere for this low magnetic field systems,
 probably depends also on the kind of accretion flow rather than 
just on the accretion rate and neutron star dipolar field strength. 

The presence of parallel tracks in the QPO frequency versus
intensity plot, seems to imply that there is more than one
parameter that determines the QPO frequency. It is expected
that one variable parameter is the accretion rate which should
determine the accretion flow and hence QPO properties. It is not
clear which is the other varying parameter which gives rise to
shifts in the correlation leading to the parallel tracks.
\cite{V01} showed that this problem can be resolved if the
QPO frequency depends on both the prompt and time-averaged value
of a single variable parameter. In this scenario, the parallel tracks
arise because the time-averaged intensity (or accretion rate) is different
for the different tracks. In the simplest version of this model, there
is no preferred intensity levels and the distinct tracks are obtained
due to observational windowing i.e. the tracks are not
connected due to  gaps in observations. However, the results obtained
in this work suggest that the parallel tracks may correspond to
definite flux levels and the gaps in the plot are not due to
under sampling. Thus the above model needs to be modified to take
into account these observations. 

The QPO phenomenon seems to be correlated to long term flux
levels on the source in a complex manner, which makes simple
theoretical interpretations difficult. The problem is enhanced
because as shown in this work the variation of the amplitude of
the QPO is affected by the incompleteness arising from the data selection
technique. Further, it has been tacitly assumed that in all the flux
states, the same QPO is being studied. A clearer idea would arise after a study of the
spectral properties of this system.  A systematic study of the spectra ( not just
the colour which is affected by detector response) of the source
at different flux levels and when the QPO is present or not, would
shed light into the phenomenon's origin. It may then be possible,
to develop and test theoretical models regarding the nature of these
systems.

\section*{Acknowledgments}

The authors thank H. C. Lee for use of the power spectra computing code
``bb''. K. S. acknowledges IUCAA for the visiting associateship and S.Shrimali, UGC-ASC for his encouragement.

\label{lastpage}


\begin{thebibliography}{}
\bibitem[Cui (2000)]{C00} Cui W., 2000, ApJ, 534, L31

\bibitem[Lee \etal (2001)]{L01} Lee H. C., Misra R., Taam R. E.,
2001, ApJ, 549, L229

\bibitem[Lee \etal (2004)]{L04} Lee W. H., Abramowicz M. A.,  Kluzniak W.,
2004, ApJ, 603, L93

\bibitem[Leahy \etal (1983)]{leahy} Leahy D. A. \etal 1983, ApJ, 266. 160

\bibitem[Mendez et al. (1999)]{M99}Mendez M.,  van der Klis M., Ford E. C.,
Wijnands R.,  van Paradijs J., 1999, ApJ, 511, L49

\bibitem[Mendez et al. (2001)]{M01}Mendez M.,  van der Klis M.,  Ford E. C., 2001, ApJ, 561, 1016

\bibitem[Miller et al. (1998)]{Mi98} Miller M.C., Lamb F.K., Psaltis D., 1998, ApJ, 508, 791

\bibitem[Osherovich \& Titarchuk (1999)]{OT99} Osherovich V., Titarchuk L., 1999, ApJ, 522, L113



\bibitem[Di Salvo \etal (2003)]{smv00} Di Salvo T., Mendez M., van~der~Klis M., 2003, { A \& A}, 406, 177

\bibitem[Stella \& Vietri (1999)]{ST99} Stella L.,  Vietri M. 1999, Phys. Rev. Lett., 82, 17

\bibitem[van Straaten et al. (2000)]{Str00} van Straaten, S., Ford, E. C., van~der~Klis M., Mendez, M., Kaaret, P. 2000, ApJ, 540, 1049.

\bibitem[van der Klis (2000)]{vdK00} van~der~Klis M. 2000, ARA\&A, 38, 717

\bibitem[van der Klis (2001)]{V01} van~der~Klis M. 2001, ApJ, 561, 943


\bibitem[Wijnands \etal (1997)] {W97}Wijnands R. A. D., \etal , 1997, ApJ, 479, L141

\bibitem[Zhang \etal (1996)] {Z96} Zhang W., Lapidus I., White N. E., Titarchuk L. 1996, ApJ, 473, L135
\end{thebibliography}
\end{document}